\documentclass[prd,11pt,aps,superscriptaddress,floatfix,showkeys,showpacs]{revtex4}
\usepackage{graphicx,amsmath,bbm,bm,array}
\usepackage[linktocpage=true,colorlinks=true,linkcolor=blue,citecolor=blue]{hyperref}

\begin{document}
\title{Shear viscosity $ \eta $ to electrical conductivity $\sigma_{el}  $ ratio for an anisotropic QGP }
\author{Lata Thakur}
\email{latathakur@prl.res.in}
\affiliation{Theory Division, Physical Research Laboratory, Navrangpura, Ahmedabad 380 009, India}
\author{P. K. Srivastava\footnote{Present address : Department of Physics, Indian Institute of Technology Ropar, Ropar 140001, India}}
\email{prasu111@gmail.com}
\affiliation{ Department of Physics, Institute of Science,
	Banaras Hindu University, Varanasi 221005, India}
\author{Guru Prakash Kadam}
\affiliation{ Department of Theoretical Physics, Tata Institute of Fundamental Research, Homi Bhabha Road, Colaba, Mumbai 400 005, India}
\author{Manu George}
\affiliation{Theory Division, Physical Research Laboratory, Navrangpura, Ahmedabad 380 009, India}
\author{Hiranmaya Mishra}
\email{hm@prl.res.in}
\affiliation{Theory Division, Physical Research Laboratory, Navrangpura, Ahmedabad 380 009, India}

\begin{abstract}
We study the transport  
properties of strongly interacting matter in the context of ultrarelativistic heavy ion collision experiments. We calculate  
the transport coefficients viz. shear viscosity ($\eta$) and electrical conductivity ($\sigma_{\rm{el}}$) 
of the quark gluon plasma phase in the presence of momentum anisotropy arising from different expansion rates of the medium in longitudinal and transverse direction.
 We solve the relativistic Boltzmann kinetic equation   
 in relaxation time approximation to calculate the shear viscosity and electrical conductivity. The calculation are performed within the quasiparticle model to estimate these transport coefficients and discuss the connection between them. 
We also compare the electrical conductivity results calculated from the quasiparticle model with the ideal case.
We compare our results with the corresponding results obtained in different lattice as well as model calculations. 
\end{abstract}


\maketitle 
\section{Introduction}
\noindent
Relativistic heavy ion collisions at the RHIC and LHC
have produced a very hot and dense pocket of nuclear matter, known as quark gluon plasma (QGP) \cite{Gyulassy:2004zy}.
Many experimental studies have been done in order to characterize the important properties of such matter produced
under extreme conditions of temperature and density. The transport coefficients for strongly interacting matter are essential theoretical
inputs for hydrodynamic evolution that are critical tools to analyze the heavy ion collision data \cite{Gale:2013da,Schenke:2011zz,Heinz:2013}. 
In relativistic hydrodynamic simulations
the shear and bulk viscosity coefficients influence various observables like the flow coefficients, the transverse momentum
distribution of produced particles. Indeed, a finite but very small shear viscosity to entropy ratio ($\eta/s$)
was necessary to explain elliptic flow data that stimulated extensive theoretical studies of this ratio  for strongly interacting 
matter.

The transport coefficient viz. shear viscosity ($ \eta $), in principle, can be estimated directly using the Kubo formulation~\cite{kubo}. However,
given that QCD is strongly coupled for energies accessible in heavy ion collision experiments, this task is complicated. Further, lattice simulations at finite chemical potentials have been challenging and are limited only to small baryon chemical potential. This has lead to attempts 
to estimate shear viscosity in various effective models~\cite{Sasaki:2008fg,Dobado:2008vt,Chakraborty:2010fr,Plumari:2012ep, Zhuang:1995uf,Wiranata:2012br, Ghosh:2014yea} involving different approximation schemes.
These include relaxation time approximations to the Boltzmann equation~\cite{Danielewicz:1984ww,Khvorostukhin:2010aj,Heckmann:2012wqa, Ghosh:2013cba,Deb:2016myz,Jaiswal:2015mxa},  Kubo formalism of evaluating equilibrium correlation functions~\cite{Defu:2005hb,Iwasaki:2007iv,Alberico:2007fu,Qin:2014dqa,Lang:2013lla,Lang:2015nca,Ghosh:2015mda},
 transport simulation of Boltzmann equation~\cite{Xu:2007ns,Marty:2013ita,Puglisi:2014sha,Plumari:2012ep}, the perturbative QCD methods~\cite{Heiselberg:1994vy,Arnold:2000dr,Arnold:2003zc,Hidaka:2008dr,Chen:2013tra, Greif:2014oia,Hattori:2016cnt,Hattori:2016lqx}, as well as lattice methods~\cite{Meyer:2007ic,Meyer:2009jp}.

Another key transport coefficient is the electrical 
conductivity ($\sigma_{el}$) of the
strongly interacting matter. This enters in the hydrodynamic evolution of quark gluon matter where charge 
relaxation also plays an important role. It is also observed that the electrical conductivity of QGP influences significantly
the soft photon production through a realistic hydrodynamic simulation~\cite{Yin:2013kya} as well as in low mass dilepton enhancement~\cite{Akamatsu:2011mw}. Further, it also suggested that the electrical conductivity can be extracted from charge dependent flow parameters 
from asymmetric heavy ion collisions~\cite{Hirono:2012rt}. 
The longitudinal static electric conductivity $\sigma_{el}$
 represents the linear response of the electrically charged particle diffusion current density $ \bf J $  to an applied external electric field $ \bf E $, i.e.,
 ${\bf J}= \sigma_{el} {\bf E } $. 
After evaluating the induced electric current one can calculate the proportionality coefficient $ \sigma_{el} $.
Electrical conductivity can be derived from the Green-Kubo formula
and is related to the correlation function for a system in thermal equilibrium, i.e., $ \sigma_{\rm{el}} =\beta V\langle \vec{J}(t=0)\cdot\vec{J}(t=0)\rangle\cdot \tau$ ~\cite{green,kubo}.
Experimentally, it has been observed that very strong electric and magnetic field is created 
in non-central heavy ion collision at the RHIC and LHC in the early stage (1-2 fm/c) of the collision~\cite{Tuchin:2013ie,Hirono:2012rt}.
The produced large electrical field affects the medium and its effect depends on the $ \sigma_{el} $ of the medium.
Electrical conductivity is responsible for the production of electric 
current in the early stage of the collision. It is of fundamental importance for the strength of chiral magnetic 
effect~\cite{Fukushima:2008xe}, a signature of CP-violation of the
strong interaction.
Recently, electric conductivity has been studied by different groups~\cite{Arnold:2000dr,Arnold:2003zc,Gupta:2004,Aarts:2007wj,Buividovich:2010tn,Ding:2010ga,Burnier:2012ts,Brandt:2012jc,Amato:2013naa,Cassing:2013iz,Steinert:2013fza,Puglisi:2014pda,Finazzo:2013efa,Mitra:2016zdw,Srivastava:2015via}.
It is related to the soft dilepton production rate ~\cite{Moore:2006qn} and the magnetic field diffusion in the medium ~\cite{Baym:1997gq,Fernandez-Fraile2006}. $ \sigma_{el} $ helps us to compare the effective cross sections of a medium's constituents among several theories, including lattice gauge theory ~\cite{Gupta:2004,Aarts:2007wj,Buividovich:2010tn,Ding:2010ga,Burnier:2012ts,Brandt:2012jc,Amato:2013naa,Aarts:2014nba,Astrakhantsev:2017nrs}, transport models ~\cite{Cassing:2013iz,Steinert:2013fza}, and Dyson-Schwinger calculations \cite{Qin:2013aaa}. 
It can also be computed on the lattice from the correlation function. Thus, the study of transport coefficients is of great interest to measure the properties of strongly interacting matter.

One of the important observations 
of HICs is that the parton system generated at the early stage of the collisions
has a strong anisotropy in momentum space due to the
different expanding rate of the longitudinal and transverse directions~\cite{Romatschke:2003ms}.
In HICs the longitudinal expansion is much faster than the transverse expansion, which causes 
the medium to become much colder in the longitudinal 
direction than the transverse direction, {\em i.e.},
$k_{\perp}\gg k_{z}\sim 1/\tau$ and a local momentum anisotropy 
appears \cite{Baier:plb502}.
Anisotropy causes the parton system produced to be
unstable with respect to the chromomagnetic plasma modes~\cite{Romatschke:2003ms} 
that facilitate one to isotropize the system~\cite{Arnold:2004ti,Mrowczynski:1993qm}.

In recent years the study of anisotropic plasma has received much interest due to the fact that the QGP, which has a local momentum-space anisotropy, is subject to the chromo-Weibel instability 
\cite{Romatschke:2003ms,Arnold:2004ti,Strickland:2007fm,Mrowczynski:2000ed,
Randrup:2003cw,Arnold:2003rq,Romatschke:2004jh,Mrowczynski:2004kv,
Rebhan:prl94,Arnold:2005vb,Rebhan:2005re,Romatschke:2005pm,Schenke:2006xu,
Schenke:2006fz,Manuel:2006hg,Bodeker:2007fw,Romatschke:2006nk,
Romatschke:2006wg,Dumitru:2005gp,Dumitru:2006pz,Rebhan:2008uj}. 
The effects of these instabilities are not very clear, but  
they are very important for the QGP evolution at the RHIC or LHC.
In recent years, the effect of anisotropy has also  been studied to investigate the properties of
 quarkonium states ~\cite{Dumitru:2007hy,Dumitru:2009ni,Burnier:2009yu,Dumitru:2009fy,Margotta:2011ta,Thakur:2012eb,Thakur:2013nia}. It will be interesting to study its effects on the properties of the QGP system.
Thus, it is important to include the momentum-space anisotropic effects in 
the calculation of transport coefficients.

In this context the ratio $(\eta/s)/(\sigma_{el}/T)$ has gained attention recently in the heavy ion phenomenology~\cite{Puglisi:2014pda}. It is quite natural to expect that QGP is a good conductor due to deconfinement of the color charges. But a small value of the ratio $\eta/s$ indicates large scattering rates that can largely damp the conductivity especially due to chargeless gluons. Our main purpose in this work is to estimate the ratio $(\eta/s)/(\sigma_{el}/T)$ for the isotropic as well as anisotropic QGP phase by solving a Boltzmann kinetic equation in relaxation time approximation (RTA). We use the quasiparticle model \cite{Peshier:2002ww,Bannur:2006ww,Srivastava:2010xa,Peshier:1999ww}, which provides a reasonable transport and thermodynamical behavior of the QGP phase.

We organize the paper as follows. In Sec. \ref{sec:Shear viscosity}, we calculate the shear viscosity and entropy density in anisotropic medium using the relativistic kinetic theory.
In Sec. \ref{sec:Electrical Conductivity}, we calculate the electrical conductivity in the anisotropic QGP medium using the Boltzmann equation in RTA.
 In Sec. \ref{sec:Quasiparticle and Bag Model}, we discuss the distribution function in the quasiparticle as well as in the ideal case. Finally, in Sec. \ref{sec:results and discussions} we discuss our results regarding shear viscosity, entropy density, and electrical conductivity. We compare our results with the lattice as well as other phenomenological calculations and give the conclusion drawn from our work.

\section{Shear viscosity and Entropy density} 
\label{sec:Shear viscosity}
The relativistic Boltzmann transport (RBT) equation has been used to calculate the shear viscosity and entropy density.
The Boltzmann transport equation for a single particle distribution function $ f(x,k) $ can be written as~\cite{Groot}
\begin{equation}
k^{\mu}\partial_{\mu} f(x,k)=C[f], 
\label{boltzmann1}
\end{equation}
where $ C[f] $ is a collision term. The shear viscosity, $ \eta $, is admissible when the equilibrium distribution $ f^{0} $ varies in space and the velocity gradient is non-zero ($ \partial_{i}u_{i}\neq 0 $).
The stress energy tensor ($ T^{\mu\nu} $) is shifted by a small amount that is proportional to this velocity gradient. 
\begin{equation}
\Delta T^{\mu\nu}=T^{\mu\nu}-T_{(0)}^{\mu\nu},
\end{equation}
where $ T_{(0)}^{\mu\nu} $ is the energy-momentum tensor for the system in local equilibrium~\cite{Hosoya:1983xm}.
\begin{equation}\label{key}
T_{(0)}^{\mu\nu}=\int \frac{d^3k}{(2\pi)^3E}k^{\mu}k^{\nu}\left\{ g_{f}f^{0}(x,k) +g_{f}\bar{f}^{0}(x,k)+g_{b}b^{0}(x,k)\right\},
\end{equation}
and  $ T^{\mu\nu} $ is
\begin{equation}
T^{\mu\nu}=\int \frac{d^3k}{(2\pi)^3E}k^{\mu}k^{\nu}\left\{ g_{f}f(x,k) +g_{f}\bar{f}(x,k)+g_{b}b(x,k)\right\};
\end{equation}
here $ f(x,k)( {\bar f(x,k)} $) and $ b(x,k) $ are the distribution functions for quarks (antiquarks) and gluons. $ g_{f} $ and $ g_{b} $ are the degeneracy factors for quarks and gluons.
Therefore, $ \Delta T^{\mu\nu} $ becomes
\begin{equation}
\Delta T^{\mu\nu}=\int \frac{d^3k}{(2\pi)^3E}k^{\mu}k^{\nu} \left\{ g_{f}\delta f(x,k) +g_{\bar{f}}\delta \bar{f}(x,k)
+ g_{b}\delta b(x,k)\right\}.
\label{deltaT}
\end{equation}
In relaxation time approximation, $ C[f] $ in Eq. (\ref{boltzmann1}) can be written as
\begin{equation}
C[f]=-\frac{k^{\mu}u_{\mu}}{\tau_{f}} (f-f^{0}), 
\end{equation}
where $ f^{0} $ is the equilibrium  distribution function for quarks. Assuming that the distribution function ($f$) is not very far from its equilibrium distribution ($ f^{0} $). Thus, $f$ can be taken as $f=f^{0}+\delta f $ and
in this approximation Eq. (\ref{boltzmann1}) becomes
\begin{equation} 
k^{\mu}\partial_{\mu} f(x,k)  = -\frac{k^{\mu}u_{\mu}}{\tau_{f}} \delta f.
\label{RBT0}
\end{equation}
Similarly, for antiquarks and gluons having equilibrium distribution functions, $\bar {f^{0}} $ and $ b^{0} $, Eq. (\ref{boltzmann1}) can be written as
\begin{eqnarray}
k^{\mu}\partial_{\mu} \bar {f}(x,k)  = -\frac{k^{\mu}u_{\mu}}{\tau_{\bar {f}}} \delta \bar {f},\\
k^{\mu}\partial_{\mu} b(x,k)  = -\frac{k^{\mu}u_{\mu}}{\tau_{b}} \delta b.
\label{RBT00}
\end{eqnarray}
Inserting $ \delta f,\delta \bar {f} $ and $ \delta b $ from Eqs. (\ref{RBT0})-(\ref{RBT00}) into 
Eq.(\ref{deltaT}), we get
\begin{eqnarray}
\Delta T^{\mu\nu}&=&-\int \frac{d^3k}{(2\pi)^3E}\frac{k^{\mu}k^{\nu}}{k.u}\bigg\{ g_{f}\tau_{{}_f}k^{\alpha}\partial_{\alpha}f(x,k)+ g_{f}\tau_{{}_{\bar{f}}}k^{\alpha}\partial_{\alpha}\bar{f}(x,k) 
+ g_{{}_b}\tau_{{}_b}k^{\alpha}\partial_{\alpha}b(x,k)\bigg\},
\end{eqnarray}
where  $ \tau_{f} ( \tau_{\bar{f}}) $ and $ \tau_{b}$ are the relaxation time for quarks (antiquarks) and gluons respectively. Momentum density, $ T^{0i} $, is small in a local Lorentz frame 
and the space-space component of energy momentum tensor  $\Delta T^{ij} $ depends linearly on the gradients of local three velocity as~\cite{Hosoya:1983xm} 
\begin{eqnarray}
\Delta T^{ij}&=&\int \frac{d^3k}{(2\pi)^3E}k^{i}k^{j}\frac{1}{T}\Bigg[ \bigg\{g_{f}\tau_{{}_f}f^{0}(1-f^{0})
+ g_{f}\tau_{{}_{\bar{f}}}\bar{f}^{0}(1-\bar{f}^{0}) 
 + g_{b}\tau_{{}_b}b^{0}(1+b^{0})\bigg\} 
 \Bigg\{ \bigg\{E{\left(\frac{\partial k}{\partial \epsilon}\right)}_{n}\nonumber\\
 &-&{\bf k}^{2}/3E\bigg\}\partial_{l}u^{l}
 -\frac{k^{k}k^{l}}{2E}W_{kl}\Bigg\}
+\bigg\{g_{f}\tau_{{}_f}f^{0}(1-f^{0})
+ g_{{\bar{f}}}\tau_{{}_{\bar{f}}}\bar{f}^{0}(1-\bar{f}^{0})\bigg\}
\times\left(\frac{\partial k}{\partial n}\right){\epsilon}\partial_{l}u^{l}\Bigg],
\end{eqnarray}
where $ \varepsilon $ and $ n $ are the energy density and  number density. The shear $ \eta $ and bulk $ \zeta $ viscosities (we do not discuss the bulk viscosity, $\zeta $) are defined as
\begin{equation}
\Delta T^{ij}=-\zeta\delta_{ij}\partial_{k}u^{k}-\eta W_{ij}\label{etaeq},
\end{equation}
where 
\begin{equation}
W_{ij}=\partial_{i}u^{j}+\partial_{j}u^{i}-\frac{2}{3}\delta_{ij}\partial_{k}u^{k}
\end{equation}
We can calculate the proportionality constant, $ \eta $,
at zero chemical potential using Eq. (\ref{etaeq}) for the isotropic medium as
\begin{eqnarray}
\eta_{iso}&=&\frac{1}{15T}\int \frac{d^3k}{(2\pi)^3} \frac{k^{4}}{E^{2}}\large\{ 2g_{f}\tau_{{}_f}f^{0}(1-f^{0})+g_{b}\tau_{{}_b}b^{0}(1+b^{0})\large\},
\end{eqnarray}
where the equilibrium distribution functions for quark, $ f^{0} $, and gluon, $ b^{0} $, at $ \mu=0 $ can be written as :
\begin{equation}
f^{0}(x,{\bf k};T)=\frac{1}{e^{\sqrt{({\bf k}^{2}+m^{2})}/T}+1}, 
\end{equation} 
and 
\begin{equation}
b^{0}(x,{\bf k};T)=\frac{1}{e^{\sqrt{({\bf k}^{2}+m^{2})}/T}-1}. 
\end{equation}
At finite chemical potential ($ \mu \neq 0 $), the 
distribution function is different for quarks and antiquarks. 
\begin{equation}
f^{0}({\bar f^{0}})=\frac{1}{e^{(E\pm\mu)/T}+1},
\end{equation}
where $ E^{2}=k^{2}+m^{2} $ and the -(+) sign is for quarks (antiquarks).\\
Shear viscosity at $ \mu\neq 0 $
\begin{eqnarray}
\eta_{iso}&=&\frac{1}{15T}\int \frac{d^3k}{(2\pi)^3} \frac{k^{4}}{E^{2}}\large\{ g_{f}\tau_{{}_f}f^{0}(1-f^{0})+g_{\bar{f}}\tau_{{}_{\bar{f}}}{\bar{f}}^{0}(1-{\bar{f}}^{0})
+g_{b}\tau_{{}_b}b^{0}(1+b^{0})\large\}.
\end{eqnarray}
As we discussed earlier we are considering the anisotropic QGP medium for our calculation. The hot QCD plasma due to expansion and non-zero viscosity, exhibits a local anisotropy in momentum space that is given by~\cite{Romatschke:2003ms} 
\begin{equation}
{\bf\tilde{k}}^{2} = \mathbf{k}^{2} + \xi(\mathbf{k}.\mathbf{\hat{n}})^{2}, 
\end{equation}
where $ \xi $ is the anisotropic parameter and  generically defined as follows~\cite{Romatschke:2003ms}:
\begin{equation}
\xi=\frac{\langle {\bf k}_{T}^{2}\rangle}{2\langle k_{L}^{2}\rangle}-1,
\end{equation}
where ${ k}_{L}$ and ${\mathbf k}_{T}$ 
are the components of momentum parallel and perpendicular to the direction of anisotropy, $ {\mathbf n} $, respectively.
The distribution function of quarks in an anisotropic system takes the following form at $\mu=0$,
\begin{equation}
f_{\rm{aniso}} (x,{\bf k};T)=\frac{1}{e^{(\sqrt{{\bf k}^{2}
			+ \xi({\bf k}.{\bf n})^{2}+ m^{2}})/T}+1}.
\label{anisof3}
\end{equation}
and the distribution function for the gluon in anisotropic medium can be written as:
\begin{equation}
b_{\rm{aniso}} (x,{\bf k};T)=\frac{1}{e^{(\sqrt{{\bf k}^{2}
			+ \xi({\bf k}.{\bf n})^{2}+ m^{2}})/T}-1},
\label{anisob3}
\end{equation}
For small $ \xi $ limit ($\xi < 1$), Eqs. (\ref{anisof3}) and (\ref{anisob3}) can be expanded as
\begin{equation}
f_{\rm{aniso}} (x,{\bf k};T)=f^{0}-\frac{\xi}{2E_{f}T}e^{E_{f}/T}{f^{0}}^{2}({\bf {k\cdot n}})^{2},
\label{anisob2}
\end{equation}
and 
\begin{equation}
b_{\rm{aniso}} (x,{\bf k};T)=b^{0}-\frac{\xi}{2E_{b}T}e^{E_{b}/T}{b^{0}}^{2}({\bf {k\cdot n}})^{2},
\label{anisof2}
\end{equation} 
where ${\bf {k\cdot n}} =k\sin\theta\sin\phi\sin\alpha+k\cos\theta\cos\alpha$. $ \alpha $ is the angle between $ {\mathbf n} $ and the z-axis.

Nonequilibrium corrections can be computed by expanding the distribution function around equilibrium \cite{Groot}.
For the anisotropic distribution function, Eq. (\ref{anisof3}), the expression for shear viscosity, $ \eta $, becomes
\begin{eqnarray}
\eta_{aniso}&=&\frac{g_{f}\tau_{{}_f}}{15T\pi^2}\int dk \frac{k^{6}}{E_{f}^{2}}\left\{ f^{0}(1-f^{0})\right\}+\frac{g_{b}\tau_{{}_b}}{30T\pi^2}\int dk \frac{k^{6}}{E_{b}^{2}}\left\{ b^{0}(1+b^{0})\right\}
-\frac{g_{f}\tau_{{}_f}}{45T\pi^2}\xi\int dk \frac{k^{8}}{E_{f}^{2}}\nonumber\\
&\times&\Bigg\{ f^{0}(1-f^{0})
\frac{1} {2E_{f}T}-\frac{(f^{0})^2}{E_{f}T}\Bigg\}
-\frac{g_{b}\tau_{{}_b}}{90T\pi^2}\xi\int dk \frac{k^{8}}{E_{b}^{2}}\Bigg\{ b^{0}(1+b^{0})\frac{1}{2E_{b}T}
+\frac{(b^{0})^2}{E_{b}T}\Bigg\}.
\end{eqnarray}
and at finite chemical potential ($ \mu\neq 0 $)
\begin{eqnarray}
\eta_{aniso}&=&\frac{1}{30T\pi^2}\int dk \frac{k^{6}}{E_{f}^{2}}\bigg\{ g_{f}\tau_{{}_f}f^{0}(1-f^{0})+g_{\bar{f}}\tau_{{}_{\bar{f}}}{\bar{f}}^{0}(1-{\bar{f}}^{0}\bigg\}
+\frac{g_{b}\tau_{{}_b}}{30T\pi^2}
\int dk \frac{k^{6}}{E_{b}^{2}}\left\{ b^{0}(1+b^{0})\right\}\nonumber\\
&-&\frac{1}{90T\pi^2}\xi
\int dk \frac{k^{8}}{E_{f}^{2}}\Bigg\{ g_{f}\tau_{{}_f}\bigg( f^{0}(1-f^{0})\frac{1} {2E_{f}T}
-\frac{(f^{0})^2}{E_{f}T}\bigg)+g_{\bar{f}}\tau_{\bar{f}}\bigg({\bar{f}}^{0}(1-{\bar{f}}^{0})\frac{1} {2E_{f}T}-\frac{({\bar{f}}^{0})^2}{E_{f}T}\bigg)\Bigg\}\nonumber\\
&-&\frac{g_{b}\tau_{{}_b}}{90T\pi^2}\xi
\int dk \frac{k^{8}}{E_{b}^{2}}
\Bigg\{ b^{0}(1+b^{0})\frac{1}{2E_{b}T}
+\frac{(b^{0})^2}{E_{b}T}\Bigg\}.\,\, \, \,\,\, \, \,
\end{eqnarray}
In kinetic theory, the entropy density for isotropic medium  at $ \mu=0 $ can be written as~\cite{Hosoya:1983xm}
\begin{eqnarray}
s_{iso}&=&-\frac{g_{f}}{\pi^{2}}\int k^{2} dk \left\{ (1-f^{0})\log(1-f^{0})+f^{0}\log f^{0} \right\}\nonumber\\
&+&\frac{g_{b}}{2\pi^2}\int k^{2} dk \left\{ (1+b^{0})\log(1+b^{0})-b^{0}
\log b^{0}\right\},
\end{eqnarray}
and at $ \mu \neq 0 $
\begin{eqnarray}
s_{iso}&=&-\frac{g_{f}}{2\pi^{2}}\int k^{2} dk  (1-f^{0})\log(1-f^{0})+f^{0}\log f^{0}+(f^{0}\rightarrow \bar{f^{0}})\nonumber\\
&+&\frac{g_{b}}{2\pi^2}\int k^{2} dk (1+b^{0})\log(1+b^{0})
-b^{0}
\log b^{0}.\, \, \,
\end{eqnarray}
For the anisotropic medium at $ \mu=0 $ we get
\begin{eqnarray}
s_{aniso}&=&-\frac{g_{f}}{\pi^{2}}\int k^{2} dk \left\{ (1-f^{0})\log(1-f^{0})+f^{0}\log f^{0} \right\}
+\frac{g_{b}}{2\pi^2}\int k^{2} dk \big\{ (1+b^{0})\log(1+b^{0})\nonumber\\
&-& b^{0}
\log b^{0}\big\}
-\xi\frac{g_{f}}{6\pi^{2}E_{f}T}\int k^{4} dk f^{0}
(1-f^{0})\log\frac{(1-f^{0})}{f^{0}}
-\xi\frac{g_{b}}{12\pi^2E_{b}T}\int k^{4} dk b^{0}\nonumber\\
&\times&(1+b^{0})\log\frac{(1+b^{0})}{b^{0}},
\end{eqnarray}
and at $ \mu \neq 0 $ as
\begin{eqnarray}
s_{aniso}&=&-\frac{g_{f}}{2\pi^{2}}\int k^{2} dk \left\{ (1-f^{0})\log(1-f^{0})+f^{0}\log f^{0} \right\}+(f^{0}\rightarrow \bar{f^{0}})\nonumber\\
&+&\frac{g_{b}}{2\pi^2}\int k^{2} dk \big\{ (1+b^{0})\log(1+b^{0})\
-b^{0}\log b^{0}\big\}
-\xi\frac{g_{f}}{6\pi^{2}E_{f}T}\int k^{4} dk f^{0}
(1-f^{0})\log\frac{(1-f^{0})}{f^{0}}\nonumber\\
&+&\xi(f^{0}\rightarrow \bar{f^{0}})
-\xi\frac{g_{b}}{12\pi^2E_{b}T}\int k^{4} dk  b^{0}
 (1+b^{0})\log\frac{(1+b^{0})}{b^{0}}
.\, \, \, \,\, \, \,\, \,
\end{eqnarray}

\section{Electrical Conductivity} 
	\label{sec:Electrical Conductivity}
The electric conductivity ($ \sigma_{el} $) represents the response of the system to an applied electric field.  According to Ohm's law $ \sigma_{el} $ can be written as 
\begin{equation}
\bf J= \sigma_{el} \bf E,
\end{equation}
 where the proportionality coefficient $ \sigma_{el} $ is the electrical conductivity. We start our calculation from the four current ($ J^{\mu} $),
\begin{eqnarray}
J^{\mu}=  \int \frac{d^{3}k}{(2 \pi)^3E} k^{\mu}\lbrace q g_{f}f(x,k)- {\bar q} g_{\bar{f}}{\bar f(x,k)}\rbrace,
 \label{current}
\end{eqnarray}
where $ q $ and ${\bar q} $ are the charge for quarks and antiquarks. 
For the case when the chemical potential is zero ($ \mu=0 $), Eq. (\ref{current}) takes the following form:
\begin{equation}
J^{\mu}= 2q_{f} g_{f} \int \frac{d^{3}k}{(2 \pi)^3E} k^{\mu}f(x,k).
\end{equation}
In the presence of some external disturbance, $ J^{\mu}=J_{0}^{\mu} + \Delta J^{\mu}$, where
\begin{equation}
\Delta J^{\mu}= 2q_{f} g_{f} \int \frac{d^{3}k}{(2 \pi)^3E} k^{\mu} \delta f(x,k).
\label{delj}
\end{equation}
One can obtain the $ \delta f(x,k) $ by using the RBT
equation as given in Sec. \ref{sec:Shear viscosity}. In the presence of the external field that is not directly related with the momentum, the RBT equation can be written in RTA as follows~ \cite{Yagi, Cercignani},
\begin{equation} 
\label{Boltzmann_eq}
k^{\mu}\partial_{\mu} f(x,k) + q F^{\alpha\beta}k_{\beta} \frac{\partial}{\partial k^{\alpha}} f(x,k) = -\frac{k^{\mu}u_{\mu}}{\tau} \delta f,
\end{equation}
where $F^{\alpha\beta}$ is the electromagnetic field strength tensor. As we are only interested in the electric field 
components of the field strength tensor ($F^{\alpha\beta}$), we take only 
$ F^{0i}=-{\bf E}~ $ and $~ F^{i0}={\bf E} $. Thus, 
the RBT equation [Eq. \ref{Boltzmann_eq}] becomes
\begin{equation}
q\left( k_0 {\bf E} \cdot \frac{\partial f^{0}}{\partial {\bf k}} + {\bf E}\cdot {\bf k} \frac{\partial f^{0}}{\partial k^0} \right) = -\frac{k^0}{\tau} \delta f.
\label{RBT1}
\end{equation}
 After solving Eq.  (\ref{RBT1}) for the anisotropic distribution function, $ f_{aniso} $ [Eq. \ref{anisof3}] 
and substituting $ \delta f $ in Eq. (\ref{delj}), 
 we obtain the expression for $\sigma_{el}$ as,
\begin{eqnarray}
\sigma_{\rm{el}}^{\rm{aniso}} (\mu_q=0) &=& \frac{1}{3\pi^2 T} \sum_f 
g_{{}_f} q_{{}_f}^{2} \int dk \frac{{\bf k}^4}{E_{f}^2}  
\tau_{{}_f} f^{0}(1-f^{0})
+\xi \frac{1}{6 \pi^2 T} \sum_f g_{{}_f} q_{{}_f}^{2} 
\int dk\frac{{\bf k}^4}{E_{f}^2}  
\tau_{{}_f}f^{0}(1-f^{0})\nonumber\\
&-&\xi \frac{1}{18 \pi^2 T} \sum_f g_{{}_f} q_{{}_f}^{2} 
\int dk\frac{{\bf k}^6}{E_{f}^2}  
\tau_{{}_f}\bigg[f^{0}(1-f^{0})
 \left(\frac{1}{E_{_{f}}^2}+\frac{1}{E_{f} T}\right)-\frac{2}{E_{f}T}(f^{0})^2\bigg].
\label{siganiso1}
\end{eqnarray}
For $ \xi=0 $, the above expression reduces to
\begin{equation}
\sigma_{\rm{el}}^{\rm{iso}} = \frac{1}{3\pi^2 T} \sum_f 
g_{{}_f} q_{{}_f}^{2} \int dk \frac{{\bf k}^4}{E_{f}^2}  
\tau_{{}_f} f^{0}(1-f^{0}).
\label{sigiso}
\end{equation}
The electrical conductivity for 
$ \mu_{q(\bar{q})} \neq 0 $ 
\begin{eqnarray}
\sigma_{\rm{el}}^{\rm{aniso}} (\mu_{q(\bar{q})} \neq 0)&=&\frac{1}{6\pi^2 T} \sum_f 
g_{{}_f} q_{{}_f}^{2} \int dk \frac{{\bf k}^4}{E_{f}^2}  \left[\tau_f f^{0}(1-f^{0})+ \tau_{\bar f}{\bar f^{0}}(1-{\bar f^{0}})\right]\nonumber\\
&+&\xi \frac{1}{12 \pi^2 T} \sum_f g_{{}_f} q_{{}_f}^{2} 
\int dk\frac{{\bf k}^4}{E_{f}^2}  
\left[\tau_f f^{0}(1-f^{0})+ \tau_{\bar f}{\bar f^{0}}(1-{\bar f^{0}})\right]\nonumber\\
&-&\xi \frac{1}{36 \pi^2 T} \sum_f g_{{}_f} q_{{}_f}^{2} 
\int dk\frac{{\bf k}^6}{E_{f}^2}  
\bigg[\left[\tau_f f^{0}(1-f^{0})+ \tau_{\bar f}{\bar f^{0}}(1-{\bar f^{0}})\right]\nonumber\\
&\times&\left(\frac{1}{E_{_{f}}^2}+\frac{1}{E_{f} T}\right)-\frac{2}{E_{f}T}(f^{0})^2
-\frac{2}{E_{f}T}(\bar f^{0})^2\bigg].
\label{siganiso}
\end{eqnarray}
For $ \xi=0 $, the above expression reduces to
\begin{eqnarray}
\sigma_{\rm{el}}^{\rm{iso}} (\mu_q\neq 0) &=&\frac{1}{6\pi^2 T} \sum_f 
g_{{}_f} q_{{}_f}^{2} \int dk \frac{{\bf k}^4}{E_{f}^2}\left[\tau_f f^{0}(1-f^{0})+ \tau_{\bar f}{\bar f^{0}}(1-{\bar f^{0}})\right]. 
\end{eqnarray}

\section{Quasiparticle Model}
\label{sec:Quasiparticle and Bag Model}
\subsection{Effective masses and relaxation times}

In the quasiparticle model, 
all the quarks (antiquarks) have both the thermal, $m_{th}$, and the bare mass, $m_{i0}$, and hence the total effective mass can be written as
~\cite{Peshier:2002ww,Bannur:2006ww,Srivastava:2010xa}
\begin{equation}
m_{i}^{2}=m_{i0}^{2}+\sqrt{2}m_{i0}m_{th,i}+m_{th,i}^{2}.
\label{mth1}
\end{equation}
The thermal mass, $m_{th}$, which arises due to the interaction of quarks (antiquarks) with the constituents of the medium, can be expressed as~\cite{Peshier:2002ww,Peshier:1999ww,Braaten:1991gm}
\begin{equation}
m_{th,i}^{2}=\frac{g^{2}(T)T^{2}}{6}\left(1+\frac{\mu_{i}^{2}}{\pi^{2}T^{2}}\right),
\label{mth}
\end{equation}
where  $g^{2}$ is the QCD running coupling constant up to two-loop order that is dependent
on both the temperature ($ T $) and chemical potential ($ \mu $)~\cite{Bannur:2006js,Zhu:2009zzi},
\begin{eqnarray}
\alpha_{S}(T)&=&\frac{g^{2}(T)}{4 \pi}=\frac{6 \pi}{\left(33-2 N_{f}\right)\ln \left(\frac{T}{\Lambda_{T}}\sqrt{1+a\frac{\mu_{q}^{2}}{T^2}}\right)}\nonumber\\
&\times&\left(1-\frac{3\left(153-19 N_f \right)}{\left(33-2 N_f\right)^2}\frac{\ln \left(2 \ln \frac{T}{\Lambda_T}\sqrt{1+a\frac{\mu_{q}^{2}}{T^2}} \right)}{\ln \left(\frac{T}{\Lambda_{T}}\sqrt{1+a\frac{\mu_{q}^{2}}{T^2}}\right) }\right),\, \, \, \, \, \, \,
\end{eqnarray}
\noindent
where  $\Lambda_{T}$ is the QCD scale parameter and the parameter $ a $ is equal to $\frac{1}{\pi^{2}}$~.\\
The $ \tau_{f} $ is the relaxation time for quarks, antiquarks, and gluons [Eqs. (\ref{sigiso}) and (\ref{siganiso1})] that can be calculated by using the following expressions in Ref.~\cite{Hosoya:1983xm} for the massless case
\begin{equation}
\tau_{q(\bar{q})}=\frac{1}{5.1 T\alpha_{s}^{2} \log \left(\frac{1}{\alpha_{s}}\right)\left(1+0.12 (2 N_{f}+1)\right)}
\end{equation}
\begin{equation}
\tau_{g}=\frac{1}{22.5 T \alpha_{s}^{2} \log \left(\frac{1}{\alpha_{s}}\right)\left(1+0.06 N_{f}\right)},
\label{tau}.
\end{equation}
Note that we have used the relaxation time for the massless case for simplicity. Our results do not change much for the massive particles case as well. Further, as shown in Ref.~\cite{Berrehrah:2013mua}, it is clear that the effect of the massive quark is small in the estimation of the scattering cross-sections. Thus, it results in a negligible effect on the relaxation time estimation. Here the results for the dissipative coefficients remain qualitatively unchanged.

In the {\em ideal case}, partons are treated as particles having rest mass only and interact weakly.
Thus, the distribution function of the ideal case contains only the rest mass term while the distribution function of the quasiparticle model (QPM) contains the rest as well as thermal mass [Eq. \ref{mth1}]. 
Here we take the rest mass of the quarks, $  m _0 =8$~MeV, for two light quarks $ u $ and $ d $ and $  m _0 =80$~MeV, for the strange quark~\cite{Srivastava:2010xa}.

\section{Results and Discussions}\label{sec:results and discussions}
\begin{figure}
\includegraphics[scale=0.45]{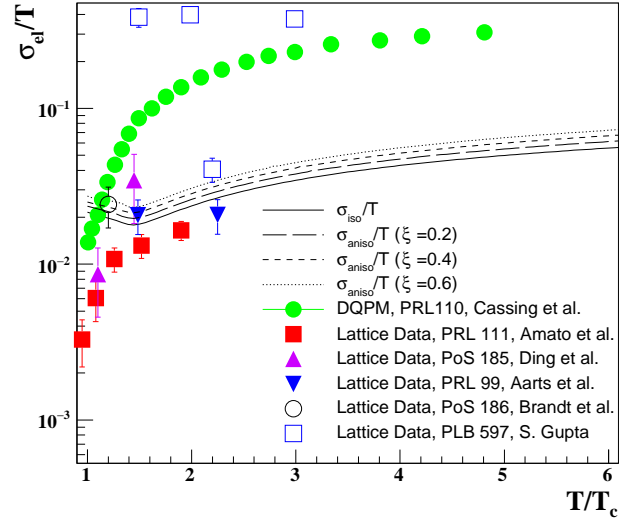}
\caption{Variation of $\sigma_{el}/T$ with respect to $T/T_c$ for isotropic (solid line) and anisotropic QGP (i.e., $ \xi=0.2,0.4,0.6 $ etc) in the present calculation. Comparison with a different lattice result is also shown.}
\label{fig1}
\end{figure}
In Fig. \ref{fig1}, we have shown the variation of the ratio of electrical conductivity to temperature ($\sigma_{\rm{el}}/T$) with respect 
to $ T/T_c $ at zero chemical potential for both the  anisotropic [Eq. \ref{siganiso1}] and isotropic [Eq. \ref{sigiso}] medium. Here we take $ T_c=180 $ MeV as the critical temperature corresponding to the quark-hadron phase transition. We found that $\sigma_{\rm{el}}^{\rm{iso}}/T$ increases monotonically with an increase in temperature. This shows that near the critical temperature, the system is  electrically less conductive than at the higher temperatures. The QCD plasma becomes opaque to transport any electrical charge at the time of phase transition. In the case of anisotropic plasma, we have observed that as the $ \xi $ increases from $0.0$ to $0.6$, the $\sigma_{\rm{el}}^{\rm{aniso}}/T$ increases for all the values of temperature. This suggests that momentum anisotropy causes the system to behave electrically more conductively.
We have compared our model results with the corresponding 
 dynamical quasiparticle model (DQPM) results (green points)~\cite{Cassing:2013iz} as well as with the data points from various lattice  calculations~\cite{Amato:2013naa,Ding:2010ga,Aarts:2007wj,Gupta:2004}. 
From Fig. \ref{fig1} we found that DQPM results overestimate the value of 
$\sigma_{\rm{el}}/T$ as compared to our model results and lattice results. Since the lattice results are distributed over a wide range, we cannot say the exact 
status of any model.

\begin{figure}
\includegraphics[scale=0.45]{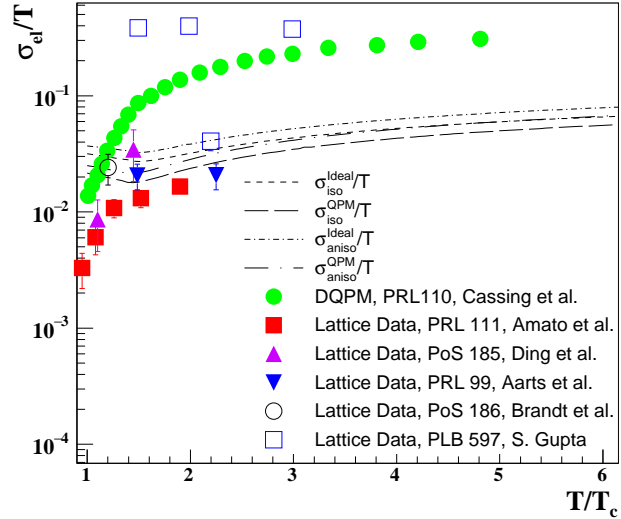}
\caption{Variation of $\sigma_{el}/T$ with respect to $T/T_c$ for anisotropic QGP ($ \xi=0.4 $) in the quasiparticle model (quarks having thermal mass) and the ideal case (no thermal mass for quarks). Different data points from the lattice are the same as in Fig. \ref{fig1}.}
\label{fig2}
\end{figure}

Figure \ref{fig2} demonstrates the comparison between the ideal case and quasiparticle model for both the isotropic and anisotropic medium. We found that $\sigma_{aniso}/T$ is more with the ideal case calculation as compared to the quasiparticle model calculation and the ratio increases with the anisotropy. This gives the possible hint to the role of thermal mass in the electrical conductivity of QCD plasma.

\begin{figure}
\includegraphics[scale=0.45]{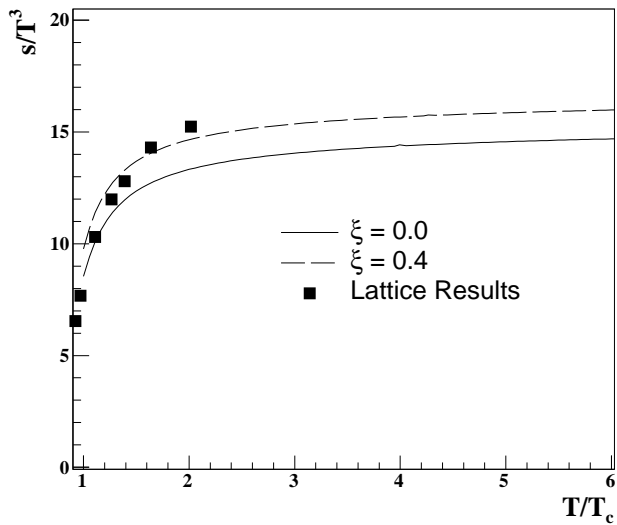}
\caption{Entropy density normalized by $T^{3}$  with respect to $T/T_c$ for isotropic (solid line) and anisotropic (dashed  line) QGP in the present calculation. Symbols represent the lattice data taken from~\cite{Borsanyi:2010cj}}.
\label{fig3}
\end{figure}
Figure \ref{fig3} shows the variation of $s/T^{3}$ with respect to $T/T_c$ at zero chemical potential. The solid line represents the QPM results for the isotropic case and the dashed line represents the anisotropic case. The data points in the figure are the lattice results taken from~\cite{Borsanyi:2010cj}. As shown in Fig. \ref{fig3} there is a smooth rise in entropy density in the vicinity of critical temperature $ T_c $ that supports a crossover type of phase transition. The increase in entropy is more in the presence of anisotropy. Here we have taken the anisotropic parameter, $ \xi=0.4 $. The plot suggests that the momentum anisotropy generates additional entropy in the system.\\

Shear viscosity is an important quantity to quantify the properties of QCD plasma. In isotropic plasma, shear viscosity has only one contribution, which comes from the collisional mode. However in anisotropic QGP, anomalous viscosity also arises due to momentum-space anisotropy along with collisional viscosity. The total viscosity of any system is dominated by the contribution that has a lower value. This anomalous viscosity may give the medium the character of a nearly perfect fluid even at moderately weak coupling.
\begin{figure}
\includegraphics[scale=0.45]{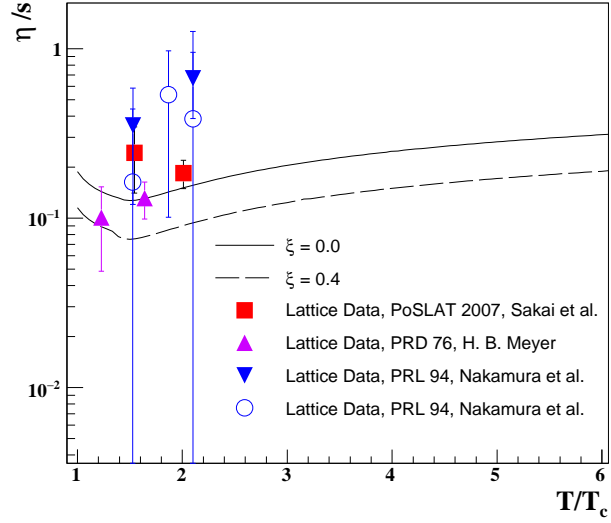}
\caption{(color online) Variation of shear viscosity to entropy ratio, $ \eta/s $,  with respect to $T/T_c$ for isotropic (solid line) and anisotropic (dashed line) QGP in the present calculation. Different lattice data results are shown by various symbols.}
\label{fig5}
\end{figure}
In Fig. \ref{fig5} we have shown the variation of shear viscosity to entropy density ratio, $ \eta/s $, with $T/T_c$ at zero chemical potential. From the figure we found that the $ \eta/s $ ratio first decreases and then increases monotonically with the increase in temperature. The $ \eta/s $ ratio decreases in the presence of anisotropy (dashed line) and keeps the same pattern as in the $ \xi=0 $ (solid line) case. Our results are in agreement with a few of the lattice results, which shows large uncertainties. From Fig. \ref{fig5}, it is clear that the collisional viscosity is high in comparison to anomalous viscosity generated due to momentum-space anisotropy. Consequently, it is actually the anomalous viscosity that makes the system behave as a perfect fluid and thus suggests that QCD plasma may not be very strongly interacting~\cite{bass}.

\begin{figure}
\includegraphics[scale=0.45]{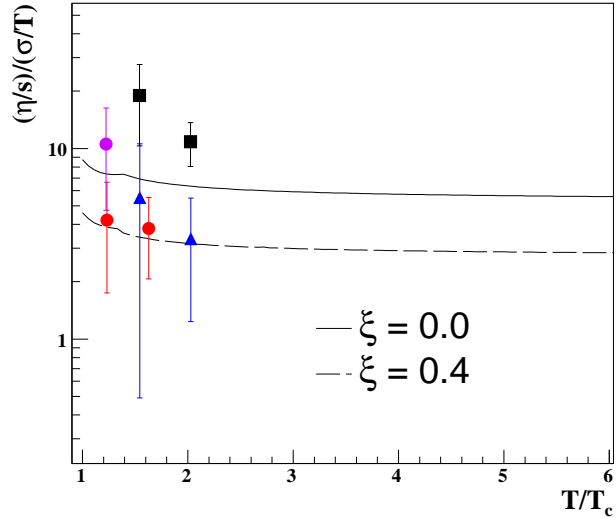}
\caption{Shear viscosity $ \eta/s $ to electrical conductivity $\sigma_{el}/T$ ratio with respect to $T/T_c$ for isotropic (solid line) and anisotropic (dashed line) QGP in the present calculation. Interpolated lattice results are taken from~\cite{Puglisi:2014pda}.}
\label{fig6}
\end{figure}
Figure \ref{fig6} shows the variation of  $(\eta/s)/(\sigma_{el}/T)$  with respect to $T/T_c$ at $\mu=0$. The solid line represents $(\eta/s)/(\sigma_{el}/T)$  for the isotropic case and the dashed line represents the anisotropic case. We have compared our quasiparticle model results with the interpolated lattice results taken from Ref.~\cite{Puglisi:2014pda}. We found that $(\eta/s)/(\sigma_{el}/T)$ starts from a large value near $ T=T_c $ and then decreases sharply with temperature and remains almost constant at higher temperatures. This suggests that the gluonic contribution in the total scattering cross-section is large near $T_{c}$ in comparison to the quark contribution and as the system departs from the phase transition point the contribution from quarks increases and starts to play a role. The ratio $(\eta/s)/(\sigma_{el}/T)$ decreases in the presence of anisotropy in the entire temperature range. As we know $\eta/s$ is effected by the contribution from gluon-gluon scattering and quark-quark scattering while $\sigma_{el}/T$ is effected only via quark-quark scattering. Thus, if $(\eta/s)/(\sigma_{el}/T)$ decreases due to anisotropy, it means that anisotropy causes either a reduction in the contribution from gluonic sector or an enhancement in the contribution from quark sector.

\begin{figure}
\includegraphics[scale=0.45]{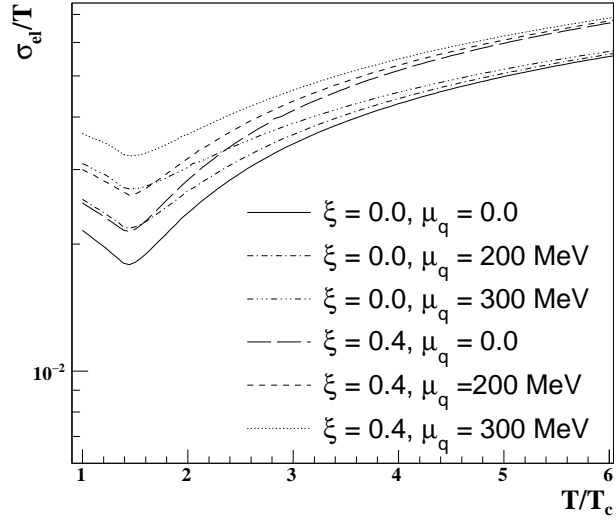}
\caption{Variation of $ \sigma_{el}/T $ with respect to $ T/T_{c} $ at finite chemical potential for both the isotropic ($ \xi=0 $) and anisotropic ($ \xi=0.4 $) case.
}.
\label{mu1}
\end{figure}

\begin{figure}
\includegraphics[scale=0.45]{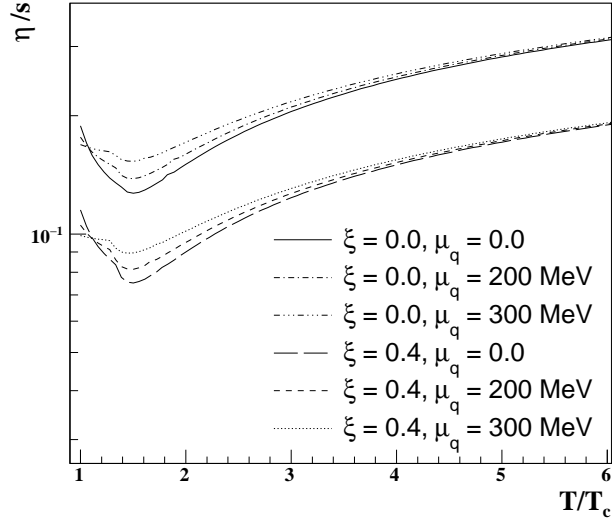}
\caption{Variation of shear viscosity to entropy ratio, $ \eta/s $,  with respect to $T/T_c$ at finite $ \mu $ for $ \xi=0 $ and $ \xi=0.4 $. }
\label{mu3}
\end{figure}

\begin{figure}
\includegraphics[scale=0.45]{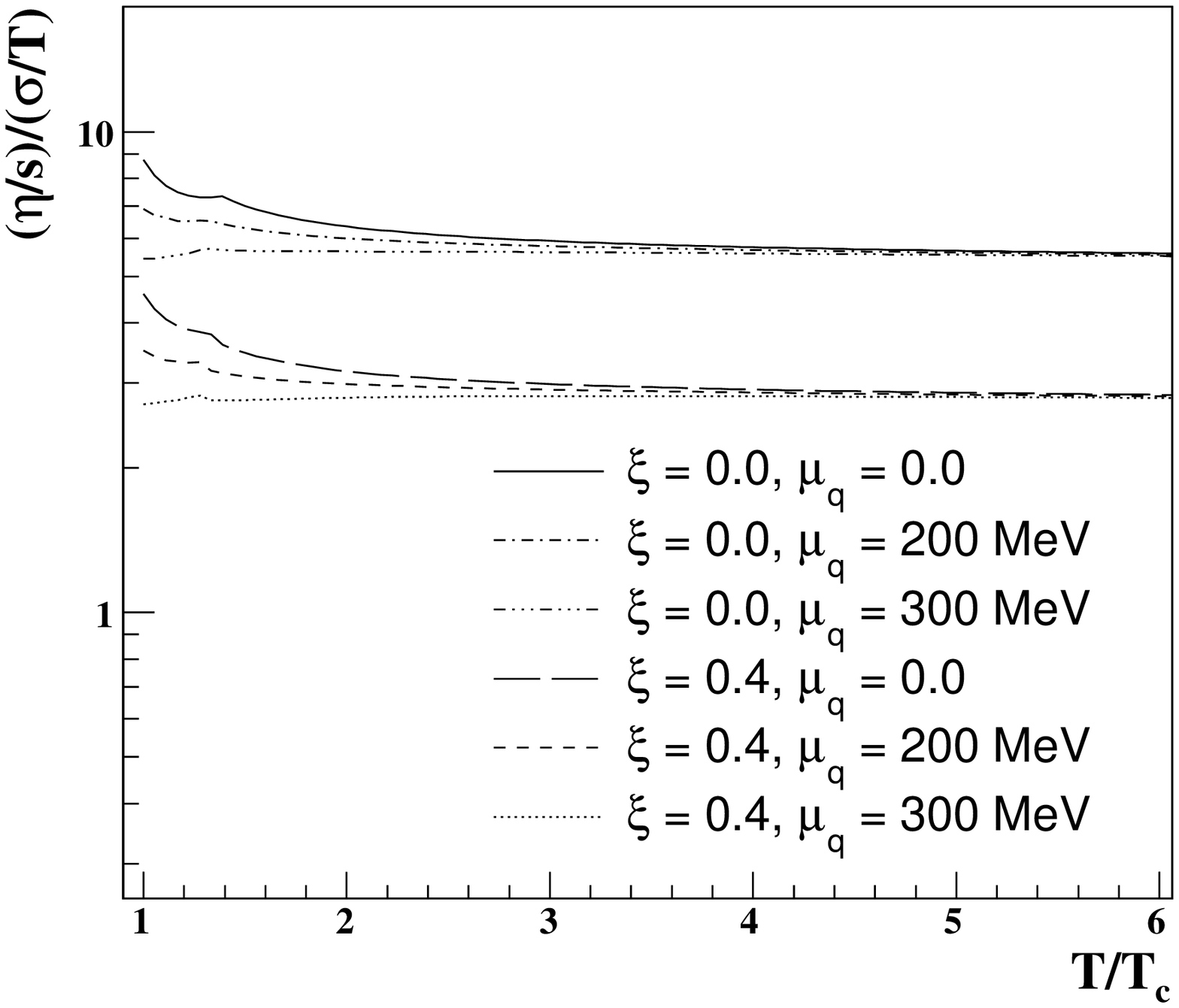}
\caption{Shear viscosity $ \eta/s $ to electrical conductivity $\sigma_{el}/T$ ratio with respect to $T/T_c$ at finite $ \mu $ for $ \xi=0 $ and $ \xi=0.4 $.}.
\label{mu4}
\end{figure}
Figure \ref{mu1} shows the variation of $ \sigma_{el}/T $ with respect to $ T/T_{c} $ at finite quark chemical potential, i.e., $ \mu = 0, 200$, and $300$ MeV for both isotropic ($ \xi=0 $) and anisotropic ($ \xi=0.4 $) cases.
From the figure we observe that the finite $ \mu $ effect is significantly large at lower temperatures as compared to higher temperatures.
 The value of $\sigma_{el}/T $ is large at finite $ \mu $  as compared to zero chemical potential and its value increases with increase in the value of $ \mu$. This significant effect at lower temperatures is due to a sizable change in distribution function of quarks at these temperatures since the ratio $\mu/T$ is significant and as the temperature increases $\mu/T$ becomes small and therefore the role of finite chemical potential diminishes on the distribution function as well as on the electrical conductivity.

Note that we have presumed a weak dependence of relaxation time on $ \mu $ and have taken $ \tau_{q(\bar{q})} $ as given in Eq.~(\ref{tau}). The $ \mu $ dependence on transport coefficients arises solely from the $ \mu $ dependence of the distribution functions. 
In Fig. \ref{mu3} we have shown the variation of $ \eta/s $  with respect to $T/T_c$ at finite $ \mu $ ({\em viz.}, $ 0,  200$, and $300$ MeV ) for $ \xi=0 $ and $ \xi=0.4 $. Similar to $ \mu=0 $ case (Fig. \ref{fig5}), we found that the $ \eta/s $ ratio first decreases and then increases monotonically with the increase in temperature at finite $ \mu $. The ratio $ \eta/s $ increases with the increase in the value of chemical potential. However, the effect of finite $ \mu$ is much less at high temperature (above $ 4T_c $). 
Figure \ref{mu4} represents the effect of finite chemical potential on the ($ \eta/s $)/($\sigma_{el}/T$) ratio. We found that the ratio decreases with the increase in chemical potential. The effect of finite $ \mu $ is more pronounced at lower temperature as compared to higher temperature.

\section{Summary}
In summary, we have studied the transport coefficients, {\em viz.}, shear viscosity ($\eta$), electrical conductivity ($\sigma_{\rm{el}}$), and thermodynamic quantity entropy density ($ s $) of the QGP phase in the presence of momentum anisotropy and discussed the connection between them.
The relativistic Boltzmann kinetic equation has been solved in RTA to calculate the $\eta$ and $\sigma_{\rm{el}}$ for the QGP phase. First we revisited the expression for shear viscosity for the isotropic medium and then derived it for the anisotropic medium by introducing the momentum anisotropy in the distribution functions of quarks,antiquarks, and gluons. Similarly, we have calculated the entropy density and electrical conductivity for the anisotropic medium.
We have shown the variation of $\sigma_{\rm{el}}/T$ with respect to $ T/T_c $ 
for both the isotropic and anisotropic medium. We found that the conductivity increases with increase in anisotropic parameter $ \xi $.  

Further, we have shown the difference arising in transport properties of QCD plasmas due to two different equations of state derived from the quasiparticle model and ideal case, respectively. We have shown the variation of entropy density with $ T/T_c $ and found a smooth rise in entropy density in the vicinity of $ T_c $ that increases in the presence of momentum anisotropy. Therefore, we can say that anisotropy generates additional entropy in the system.  
We have also shown the effect of anisotropy on the $ \eta/s $ ratio (Fig.\ref{fig5}) and found that it decreases with increase in anisotropy. From this result one may infer that anomalous viscosity that arises due to momentum anisotropy makes the system behave as a perfect fluid. Our results are in agreement with a few of the lattice results, which show large uncertainties.
We have discussed the 
variation of  $(\eta/s)/(\sigma_{el}/T)$  with respect to $T/T_c$. We found that quark contribution in the total scattering cross-section is less near $T_{c}$ in comparison to gluon contribution and at higher temperature quark contribution increases and plays a significant role. The presence of anisotropy results in a decrease in the ratio $(\eta/s)/(\sigma_{el}/T)$ in the entire temperature range and thus provides a hint regarding the change in the contribution of the gluonic sector.

Finally, we have shown the effect of finite chemical potential, i.e., $ \mu=200$, and $300 $ MeV on $ \sigma_{el}/T $, $ \eta/s $, and $(\eta/s)/(\sigma_{el}/T)$ for both the isotropic and anisotropic cases. 
Within the quasiparticle approximations the transport coefficients turn out to be larger at finite $ \mu $ as compared to their value at vanishing chemical potential.
The finite $ \mu $ effect is more significant at lower temperature as compared to higher temperature due to the sizable change in the distribution function at lower temperature as compared to higher temperature. 


\end{document}